\documentclass[twocolumn,10pt]{article}

\usepackage[utf8]{inputenc}
\usepackage[english]{babel}
\usepackage{amsmath,amssymb,amsfonts} 
\usepackage{graphicx}                  
\usepackage{authblk}                  
\usepackage{hyperref}
\usepackage{sectsty}
\allsectionsfont{\raggedright}

\usepackage[left=2cm,right=2cm,top=2cm,bottom=2cm]{geometry}

\title{\Large\textbf{On the Stability of the Polar Phase of Superfluid $^3$He in Nematic Aerogels }}

\author{\textbf{E.~V.~Surovtsev}\thanks{e-mail: e.v.surovtsev@gmail.com}}

\affil{\small P.~L.~Kapitza Institute for Physical Problems, Russian Academy of Sciences, Kosygina 2, 119334 Moscow, Russia}

\date{}

\begin{document}
	
	\twocolumn[
	\begin{@twocolumnfalse}
		\maketitle
		
		\begin{abstract}
			\noindent We study the stability of the solution corresponding to the polar phase of 
			superfluid $^3\text{He}$ in nematic aerogels with respect to perturbations induced 
			by magnetic impurity scattering. We show that when the perturbation 
			simultaneously breaks axial and time-reversal symmetry, the 
			polar-distorted $A$~phase is realized with the highest transition temperature. 
			Within the proposed model of a $p$-wave scattering potential, which accounts 
			for the effective spin-orbit interaction between the orbital angular momentum 
			of a scattering quasiparticle and the magnetic moment texture of the scatterer, 
			we solve the eigenvalue and eigenvector problem for the Gor'kov 
			self-consistency equation. Finally, we discuss how our results relate to the 
			available experimental data.
		\end{abstract}
		\vspace{0.6cm} 
	\end{@twocolumnfalse}
	]
	
\section{Introduction}

In pure superfluid $^3\text{He}$, triplet pairing with $L=1$ occurs, where the 
order parameter is a complex $3\times3$ matrix. In this simplest case, all 
order parameter components share the same transition temperature, and the 
superfluid phase that minimizes the free energy is realized. Anisotropic 
aerogels lift this temperature degeneracy, enabling phases that cannot exist 
in the pure bulk liquid. This occurs because the superfluid transition 
temperatures for the order parameter components corresponding to different 
projections of the orbital angular momentum become distinct. Nematic aerogels 
represent one type of aerogel available for experimental studies. This material 
consists of thin, co-aligned strands that form an elastic framework. As 
previously shown theoretically, for axially symmetric nematic aerogels with 
scattering that preserves the longitudinal momentum component (specular scattering), 
the polar phase—i.e., the superfluid state with a zero projection of the orbital 
angular momentum onto the anisotropy axis—exhibits the highest transition 
temperature \cite{Fomin_2018}. Moreover, under this scattering mechanism, 
the transition temperature remains unchanged and equals that of pure $^3\text{He}$, 
meaning that Anderson's theorem holds for the polar phase of superfluid $^3\text{He}$ \cite{Anderson}. 
Although diffuse scattering violates Anderson's theorem and suppresses the transition 
temperature, the polar phase still maintains the highest transition temperature 
among all phases \cite{Aoyama, Ikeda_2020, Sauls_2021}. In other words, the polar 
phase is stable against this type of perturbation. This stability is further 
supported by the excitation spectrum in momentum space, which features a 
topologically protected line of nodes along the equator of the Fermi surface 
(the Berry phase changes by $\pi$ when looping around the nodal line). Notably, 
this line of nodes possesses axial symmetry and is invariant under time reversal. 
In contrast to the polar phase, the $A$~phase of superfluid $^3\text{He}$ 
hosts two singular points in momentum space where the gap vanishes (Weyl points). 
These point-like features in momentum space are topologically protected 
with chiralities of $\pm 1$. A detailed symmetry analysis of the phases 
realized in nematic aerogels can be found in Ref.~\cite{Eltsov_chin}.The transformation of a line of nodes into two point-like features during a 
second-order phase transition was analyzed in Ref.~\cite{Volovik_Nis}. 
Notably, this transition spontaneously breaks time-reversal symmetry, 
since the projection of the orbital angular momentum in the $A$~phase can be $\pm1$. 
One can expect that introducing a perturbation that breaks both axial and 
time-reversal symmetries will also destroy the line of nodes of the polar phase, 
potentially transforming it into two point-like features located on the equator. 
The phase arising under such perturbations is called the polar-distorted $A$~phase. 
Initially, it was suggested that this specific state was observed in the pioneering 
experiments on $^3${He} in nematic aerogels \cite{Dmitriev_2012, Dmitriev_2014_squeezed}. 
The order parameter of the polar-distorted $A$~phase can be written as:
\begin{eqnarray}
A_{\mu j}= {\Delta}\cdot\hat{d}_{\mu}\Big[\cos\frac{\alpha}{2}\cdot\hat{m}_j+i\sin\frac{\alpha}{2}\cdot \hat{n}_j\Big],
\end{eqnarray}	
where $\alpha \in [0, \pi/2]$, with $\alpha = 0$ corresponding to the polar 
phase and $\alpha = \pi/2$ to the pure $A$~phase. In subsequent experiments, 
high-symmetry aerogels—such as Nafen and aerogels with mullite strands—were used. 
When the strands of these aerogels were pre-plated with solid $^4\text{He}$, 
a state with $\alpha = 0$ was observed with high experimental precision 
near the superfluid transition at all pressures \cite{Dmitriev_polar, Dmitriev_obzor, Eltsov_2023}. 
In particular, Ref.~\cite{Eltsov_2023} demonstrated that the low-temperature 
behavior of the energy gap directly signals the presence of a line of nodes 
in the excitation spectrum. In all the above studies 
\cite{Dmitriev_2012, Dmitriev_2014_squeezed, Dmitriev_polar, Dmitriev_obzor, Eltsov_2023}, 
the nematic aerogel was pre-plated with several $^4${He} monolayers to screen 
the Van der Waals interaction between the aerogel strands and $^3${He}, 
thereby preventing the formation of a solid paramagnetic film on the surface. 
In the absence of this pre-plating, the solid paramagnetic $^3${He} adsorbed onto 
the nematic aerogel surface completely suppresses the stability region of the polar phase against the formation of the $A$~phase \cite{Dmitriev_magnetic}. 
Crucially, experiments \cite{Dmitriev_magnetic} in Nafen-72 and in 
mullite aerogels \cite{Dmitriev_Eltsov} yielded a highly intriguing result: 
under partial $^4\text{He}$ coverage—where only a small amount of $^3\text{He}$ 
is present on the strands—there are strong indications that the system undergoes a transition from the normal state into a polar-distorted $A$~phase with a finite degree of distortion. Earlier theoretical 
work showed that the exchange interaction between solid and liquid $^3\text{He}$ 
can shrink the temperature range where the polar phase exists, 
effectively driving the system toward isotropy \cite{Mineev_2018}.
The scenario proposed in that work can explain the transition into the undistorted $A$~phase; however, the realization of the polar-distorted $A$~phase goes beyond the scope of that model. The purpose of the present work is to show that accounting for the effective 
spin-orbit interaction during the scattering of liquid $^3\text{He}$ quasiparticles 
by the magnetic texture of solid $^3\text{He}$ explains the instability of the polar phase 
against the formation of the polar-distorted $A$~phase. Unlike previously 
considered magnetic scattering models, the proposed mechanism simultaneously 
breaks both axial and time-reversal symmetries.

In this work, we consider a microscopic mechanism associated with 
quasiparticle scattering on aerogel strands that leads to the instability 
of the polar phase against the formation of a polar-distorted $A$~phase. 
As noted, this requires the ensemble-averaged impurity potential to break 
both axial and time-reversal symmetries. Consequently, the potential must 
mix states with orbital angular momentum projections $m=0$ and $m=\pm1$, 
which corresponds to $p$-wave disorder. Recently, the effect of $d$-wave 
disorder on the thermodynamic properties of $d_{x^2-y^2}$-wave superconductors 
was investigated \cite{Wang}. In contrast to that study, where the primary effect 
is due to the emerging anisotropy of the density of states without altering 
the form of the order parameter, we show that even with an isotropic density 
of states in the normal phase, $p$-wave disorder can stabilize a novel 
superfluid anisotropic (chiral) phase.

The most relevant perturbation is associated, on one hand, with the 
aerogel inhomogeneity and, on the other, with the solid $^3\text{He}$ magnetic 
layer on the aerogel surface. The primary physical reason underlying the 
aerogel inhomogeneity is that the strands in a real aerogel are not perfectly 
co-aligned, which in turn leads to their mutual intersections. In addition, 
the strands themselves possess a finite curvature and a rough surface, further 
enhancing the inhomogeneity of the system. The time-reversal symmetry breaking 
arises from magnetic scattering by the paramagnetic layer of solid $^3\text{He}$. 
To link these two types of symmetry breaking (spatial and temporal), one must 
assume the existence of ferromagnetic-type correlations in the magnetic layer, 
as well as that the weak dipole-dipole interaction between the $^3\text{He}$ spins 
in the solid surface layer leads to an easy-plane ordering (where spins predominantly 
lie in the plane of the aerogel surface). Ferromagnetic magnetic ordering has been 
observed in a two-dimensional solid layer of $^3\text{He}$ on a graphite substrate 
for 2--3 adsorbed monolayers \cite{Bozler_2003}. Extrapolating these results to 
the aerogel case, one can assume that the magnetization texture is determined 
by the aerogel inhomogeneity (strand intersections, curvature, and surface roughness). 
As shown in the Appendix, quasiparticle scattering by a spatially inhomogeneous 
magnetic layer leads, in the adiabatic approximation, to an effective spin-orbit interaction.

The paper is organized as follows. In Sec.~\ref{sec:algorithm}, we present the 
algorithm for solving the Gor'kov self-consistency equation for $p$-wave 
pairing in the presence of anisotropic impurities. In Sec.~\ref{sec:model}, 
we introduce the aerogel model that features the abovementioned symmetry-breaking 
properties. Sections~\ref{sec:greens} and \ref{sec:vertex} describe the properties 
of the Green's function and the vertex part under $p$-wave disorder. 
In Sec.~\ref{sec:eigen}, we find the eigenvalues and eigenvectors of the self-consistency 
equation. Section~\ref{sec:results} is devoted to the analysis of the obtained results. 
Finally, the Appendix provides a qualitative derivation of the effective 
interaction potential, including the spin-orbit term.

\section{Self-consistency equation}

\label{sec:algorithm}
The original matrix self-consistency equation for spin-triplet Cooper pairing 
in the $p$-wave channel is expressed in terms of the anomalous
Green's function $\hat{F}_{\alpha\beta}(\mathbf{k}, \omega_n)$ and the effective 
attractive BCS potential
$V_{\alpha\beta;\gamma\varphi}(\mathbf{k}, \mathbf{k}_1) = 3\lambda\cdot (\hat{\mathbf{k}} \cdot \hat{\mathbf{k}}_1) \cdot \frac{1}{2}\left(\delta_{\alpha\gamma}\delta_{\beta\varphi}+\delta_{\alpha\varphi}\delta_{\beta\gamma}\right)$ \cite{Mineev_Samohin}:
\begin{equation}\label{eq:self_cons_matrix}
\Delta_{\alpha\beta}(\mathbf{k}) = -T \sum_{\omega_n} \int \frac{d^3 k_1}{(2\pi)^3}  V_{\beta\alpha;\gamma\varphi}(\mathbf{k}, \mathbf{k}_1) F_{\gamma\varphi}(\mathbf{k}_1, \omega_n),
\end{equation}
where $\Delta_{\alpha\beta}(\mathbf{k})$ is the anisotropic triplet gap, 
$\lambda$ is the coupling constant, and $\omega_n = \pi T(2n+1)$ with 
$n = 0, \pm 1, \dots$ are the Matsubara frequencies, assuming a spatially 
homogeneous case. Linearizing with respect to $\Delta$, the anomalous Green's 
function can be obtained from the Gor'kov equations as:
\begin{eqnarray}
\label{F_1}
F_{\alpha\beta}(\mathbf{k},\omega_n)\approx G_{\alpha\mu}(\mathbf{k}, \omega_n) \tilde{\Delta}_{\mu\nu}(\mathbf{k})G_{\beta\nu}(-\mathbf{k}, -\omega_n),
\end{eqnarray}
where $G_{\alpha\beta}(\mathbf{k}, \omega_n) = \delta_{\alpha\beta} G(\mathbf{k}, \omega_n)$ 
is the Green's function of the normal isotropic state (defined below), 
and $\tilde{\Delta}_{\alpha\beta}(\mathbf{k})$ is the gap renormalized by 
impurity scattering. Next, we use the standard representation of the triplet 
gap in terms of the macroscopic order parameter $A_{\mu j}$:

\begin{eqnarray}
\Delta_{\alpha\beta}(\mathbf{k}) = A_{\mu j} \hat{k}_j (\sigma_\mu i\sigma_y)_{\alpha\beta}.
\end{eqnarray}
Similarly, we express the renormalized gap by using the definition of the vertex part:
\begin{eqnarray}
\tilde{\Delta}_{\alpha\beta}(\mathbf{k})= A_{\mu j} \Gamma_j^{\mu\nu}(\mathbf{k}) (\sigma_\nu i\sigma_y)_{\alpha\beta},\label{Delta_t}	
\end{eqnarray}
where the superscripts in the vertex part are introduced solely for brevity.

Substituting Eqs.~(\ref{F_1}) and (\ref{Delta_t}) into the matrix equation~(\ref{eq:self_cons_matrix}) and projecting onto the triplet Pauli matrix basis, we obtain the self-consistency equation for the order parameter $A_{\mu j}$:
\begin{eqnarray}\label{eq:self_cons_tensor}
A_{\mu i} \hat{k}_i = -3\lambda T \sum_{\omega_n} \int \frac{d^3 k_1}{(2\pi)^3} (\hat{\mathbf{k}} \cdot \hat{\mathbf{k}}_1) G(\mathbf{k}_1,\omega_n)\times\nonumber\\ G^{(0)}(-\mathbf{k}_1,-\omega_n) A_{\nu j} \Gamma_j^{\mu\nu}(\mathbf{k}_1, \omega_n).
\end{eqnarray}
Finally, projecting the equations onto the corresponding orbital directions yields a system of linear homogeneous equations for the components of the order parameter matrix:
\begin{eqnarray}
\label{Eq}
{A}_{\mu j}=\Lambda_{jl}^{\mu\nu}{A}_{\nu l},
\end{eqnarray}
where the matrix $\Lambda_{jl}^{\mu\nu}$ is given by the expression:
\begin{eqnarray}
\Lambda_{jl}^{\mu\nu}= -3\lambda T \sum_{\omega_n} \int \frac{d^3 k_1}{(2\pi)^3} (\hat{k}_1)_j\Gamma_l^{\mu\nu}(\mathbf{k}_1, \omega_n)\times\nonumber\\ G(\mathbf{k}_1,\omega_n) G(-\mathbf{k}_1,-\omega_n). 
\end{eqnarray}
Equation~(\ref{Eq}) is essentially an eigenvalue and eigenvector problem for the matrix $\Lambda_{jl}^{\mu\nu}$ \cite{Fomin_2021}. In pure $^3$He, the matrix $\Lambda_{jl}^{\mu\nu}$ is trivial with respect to all indices, implying that the transition temperature is identical for all order parameter components. Potential scattering in nematic aerogel lifts the degeneracy in the orbital subspace of the order parameter. In this case, the axial symmetry of the scattering potential implies that the matrix $\Lambda_{jl}^{\mu\nu}$ becomes diagonal in the basis where the $z$ axis is directed along the aerogel anisotropy axis. Below, we show that magnetic scattering breaks both axial and time-reversal symmetries. Consequently, the matrix $\Lambda_{jl}^{\mu\nu}$ is no longer diagonal in the orbital subspace within the specified basis, requiring one to find the correct eigenfunctions (phases) in this scenario. Thus, the problem reduces to determining the structure of the matrix $\Lambda_{jl}^{\mu\nu}$.

The non-trivial form of the matrix $\Lambda_{jl}^{\mu\nu}$ follows directly from that of the vertex part. In the ladder approximation, the vertex tensor $\Gamma_j^{\mu\nu}$ satisfies the equation:
\begin{eqnarray}\label{vertex}
\Gamma_j^{\mu\nu}(\mathbf{k}) = \delta^{\mu\nu} \hat{k}_j + n_s \int \frac{d^3 k_1}{(2\pi)^3} G(\mathbf{k_1},\omega_n)\times\nonumber\\ G(-\mathbf{k_1},-\omega_n)  \mathcal{K}^{\nu\lambda}(\mathbf{k}, \mathbf{k}_1) \Gamma_j^{\mu\lambda}(\mathbf{k}_1),
\end{eqnarray}
where $n_s$ is the concentration of scattering centers, and the two-particle spin-scattering kernel $\mathcal{K}^{\nu\lambda}$ in the Born approximation is given by:
\begin{eqnarray}
\mathcal{K}^{\nu\lambda}(\mathbf{k}, \mathbf{k}_1) = \frac{1}{2} \text{Tr} \left[\right.\langle (\hat{\sigma}_\nu i \hat{\sigma}_y)^\dagger \cdot \hat{u}(\mathbf{k}, \mathbf{k}_1) \times\nonumber\\ (\hat{\sigma}_\lambda i \hat{\sigma}_y) \cdot \hat{u}^T(-\mathbf{k}, -\mathbf{k}_1) \rangle\left.\right],
\end{eqnarray}
here $\hat{u}(\mathbf{k}, \mathbf{k}_1)$ is the matrix element of the scattering potential for a single impurity, and the angular brackets $\langle...\rangle$ denote averaging over the orientations of the anisotropy axes of an individual impurity. The single-impurity scattering potential matrix can be expanded in the Pauli matrix basis:
\begin{eqnarray}
\hat{u}(\mathbf{k}, \mathbf{k}_1)=u_0(\mathbf{k}, \mathbf{k}_1)\cdot\hat{\sigma}_0+(m_{\lambda}\cdot\hat{\sigma}_{\lambda})u_m(\mathbf{k}, \mathbf{k}_1),
\end{eqnarray}
where $\mathbf{m}$ is a unit vector in spin space.
Substituting the potential into the expression for the kernel and averaging over disorder yields a structure that is trivial with respect to the spin indices:
\begin{gather}
\mathcal{K}^{\mu\nu}(\mathbf{k}, \mathbf{k}_1)=\mathcal{K}(\mathbf{k}, \mathbf{k}_1)\cdot\delta^{\mu\nu}=\nonumber\\
\delta^{\mu\nu}\cdot\Big(u_0(\mathbf{k}, \mathbf{k}_1)\cdot u_0(-\mathbf{k}, -\mathbf{k}_1)+\nonumber\\
\frac{1}{3} u_m(\mathbf{k}, \mathbf{k}_1)\cdot u_m(-\mathbf{k}, -\mathbf{k}_1)\Big).
\end{gather}
In deriving the latter expression, we have assumed an isotropic distribution of the vector $\mathbf{m}$, such that $\langle m_{\mu}m_{\nu}\rangle=\frac{1}{3}\delta^{\mu\nu}$. Consequently, the spin structure of the vertex part can be omitted, leaving only the orbital index:
\begin{eqnarray}
	\Gamma_j(\mathbf{k}) = \hat{k}_j + n_s \int \frac{d^3 k_1}{(2\pi)^3} G(\mathbf{k_1},\omega_n)\times\nonumber\\ G(-\mathbf{k_1},-\omega_n) \mathcal{K}(\mathbf{k}, \mathbf{k}_1) \Gamma_j(\mathbf{k}_1).
	\label{Gamma_0}
\end{eqnarray}
If the scattering preserves time-reversal symmetry, then $u(-\mathbf{k}, -\mathbf{k}_1)=u^*(\mathbf{k},\mathbf{k}_1)$ and $G(-\mathbf{k_1},-\omega_n)= \Big[G(\mathbf{k_1},\omega_n)\Big]^*$. Thus, provided that the scattering potential does not break time-reversal symmetry, the kernel of Eq.~(\ref{Gamma_0}) is a real-valued function. Since the matrix $\hat{\Lambda}$ determines the free energy quadratic form with respect to the order parameter, it is Hermitian. It follows immediately that, in the scenario under consideration, the matrix $\Lambda_{jl}^{\mu\nu}$ is a real symmetric matrix, whose eigenvectors can be chosen to be real. If the state with the maximum transition temperature is non-degenerate, its eigenvector represents a polar phase. Consequently, in order to obtain a Hermitian matrix $\Lambda_{jl}^{\mu\nu}$ with complex off-diagonal elements, the scattering process must break time-reversal symmetry. A complex vector (such as the polar-distorted A-phase) can serve as a solution for a non-degenerate eigenvalue only in this case. The above analysis is strictly valid within the Born approximation, which is well-suited for the case of weak impurities. Otherwise, the vertex-part equation would require the $T$ matrix instead of the Fourier transform of the potential, which is generally non-Hermitian. Below, we focus on the Born approximation and the case of time-reversal symmetry breaking.

\section {Microscopic Model of the Scattering Potential}
\label{sec:model}

The primary effect investigated in this work arises from the effective spin-orbit interaction. Therefore, it is necessary to discuss the possible microscopic origins of this interaction. As noted in the Introduction, strands pre-plated with a paramagnetic layer of solid $^3$He featuring intralayer ferromagnetic correlations generate a strong exchange magnetic field near them. The curvature of the strands, their intersections, and surface roughness cause the magnetization direction to vary smoothly in space on scales larger than the interatomic distance. Consequently, in addition to the local magnetization direction $\mathbf{m}(\mathbf{r})$, the scattering amplitude acquires terms linear in another vector: the Berry curvature vector induced by the magnetic texture, which is defined as $H_i=e_{ijk}e_{lnp}m_l\nabla_jm_n\nabla_km_p$ (see Appendix). In other words, a system of correlated magnetic impurities on a spatially inhomogeneous surface cannot be characterized solely by the average magnetization. After averaging over disorder, the system properties will also depend on higher-order correlators, with the Berry curvature vector being one of them. Since the magnetization and the Berry curvature represent correlators of different orders, both in terms of the local magnetization vector power and its spatial derivatives, they are linearly independent. Accordingly, a net non-zero Berry curvature can strictly realize in a given macroscopic region of the sample even at zero average macroscopic magnetization within the same region, due to the curvature (chirality) of the magnetization texture near strand intersections and other inhomogeneities.

In this work, we consider the following aerogel model that describes anisotropic scattering by its strands, taking into account the simultaneous breaking of axial and time-reversal symmetries. We model the nematic aerogel as a system of point-like strands aligned along the z axis, decorated with anisotropically scattering beads that simulate the local curvature of the magnetic scattering field. In fact, this approach treats a magnetically perturbed model of highly anisotropic aerogel, which was previously proposed in Ref.~\cite{Fomin_2018}. In the coordinate representation, the effective interaction operator between a quasiparticle and the strand system is written as:
\begin{eqnarray}
U_{\alpha\beta}({\mathbf{r}})=\sum\limits_{a=1}^{N_s}\Big[\delta(\boldsymbol{\rho}-\boldsymbol{\rho}_a)\cdot g_0\cdot(\sigma_0)_{\alpha\beta}+\nonumber\\
\sum\limits_{b=1}^{N_b}v_{\alpha\beta}(\boldsymbol{\rho}-\boldsymbol{\rho}_a,z-z_{ab})
\Big],
\end{eqnarray}		
where $N_s$ is the number of strands, $N_b$ is the number of beads per strand, $\mathbf{r}=(\boldsymbol{\rho},z)$, $g_0$ is the scattering amplitude of an ideal specularly scattering strand, $v_{\alpha\beta}(\mathbf{r})$ is the three-dimensional anisotropic scattering potential of a bead, and $(\sigma_0)_{\alpha\beta}$ is the $2\times2$ identity matrix. For further analysis, we also introduce the strand concentration $n_s$ and the concentration of beads per strand $n_b$.
The kernel of the interaction operator with a single magnetic bead, taking into account the effective spin-orbit interaction, is given by:
\begin{eqnarray}
[v_{Iso+SO}^{(1)}]_{\alpha\beta}({\mathbf{r}},{\mathbf{r}}^{'})=\nonumber\\
\delta({\mathbf{r}})\cdot\delta({\mathbf{r}}-{\mathbf{r}}^{'})\cdot[f_{0p}(\sigma_0)_{\alpha\beta}+f_{0e}\cdot m_{\mu}(\sigma_{\mu})_{\alpha\beta}]+\label{SO}\\
\frac{1}{2}\Big[i\hbar e_{ijk} H_ir_j\frac{\partial}{\partial r_k^{'}}\delta({\mathbf{r}}-{\mathbf{r}}^{'})+h.c.\Big]\times\nonumber\\
\times[f_{1e}\cdot m_{\mu}(\sigma_\mu)_{\alpha\beta}+f_{1p}(\sigma_0)_{\alpha\beta}],\nonumber
\end{eqnarray}
where the first term describes the isotropic contribution to the potential scattering (with amplitude $f_{0p}$), the second term corresponds to the standard isotropic exchange interaction (with amplitude $f_{0e}$), and the third and fourth terms represent the anisotropic $p$-wave channel associated with the effective spin-orbit interaction. Note that within the proposed model, the bead is treated as a point-like object; however, its effective magnetization and Berry curvature vector arise from averaging over the scattering surface of the initial inhomogeneous region on the aerogel strand, which yields an effectively non-local interaction. The origin, as well as a qualitative derivation of the second term in Eq.~(\ref{SO}) (proportional to $f_{1e}$), are discussed in the Appendix. The term proportional to $f_{1p}$ is introduced for completeness.
The Fourier transform of the total potential is given by:
\begin{eqnarray}
u_{\alpha\beta}(\mathbf{k},\mathbf{k}^{'})=2\pi (\sigma_0)_{\alpha\beta} g_0\delta(k_z-k_z^{'})\sum\limits_{a=1}^{N_s}e^{-i\boldsymbol{\rho}_a(\boldsymbol{\kappa}-\boldsymbol{\kappa}^{'})}+\nonumber\\
\sum\limits_{a=1}^{N_s}\sum\limits_{b=1}^{N_b}\Big[f_{0p}(\sigma_0)_{\alpha\beta}+f_{0e}\cdot m_{\mu}^{(a,b)}(\sigma_{\mu})_{\alpha\beta}\Big]\times\nonumber\\
\label{pot}
e^{-i\boldsymbol{\rho}_a(\boldsymbol{\kappa}-\boldsymbol{\kappa}^{'})}e^{-iz_{ab}(k_z-k_z^{'})}+\\
i\cdot (\hat{k}_l\hat{k}_m^{'}-\hat{k}_m\hat{k}_l^{'})\cdot  \sum\limits_{a=1}^{N_s}\sum\limits_{b=1}^{N_b}\Big[f_{1e}\cdot m_{\mu}(\sigma_\mu)_{\alpha\beta}+\nonumber\\f_{1p}(\sigma_0)_{\alpha\beta}\Big]\cdot
h_{lm}^{(a,b)}(k,k^{'})\cdot e^{-i\boldsymbol{\rho}_a(\boldsymbol{\kappa}-\boldsymbol{\kappa}^{'})}e^{-iz_{ab}(k_z-k_z^{'})},\nonumber
\end{eqnarray}
here ${h}_{lm}^{(a,b)}= e_{lmn}\cdot \frac{H_n^{(a,b)}}{|H^{(a,b)}|}$ is the Berry curvature tensor of the $b$-th bead on the $a$-th strand, which is an antisymmetric tensor that changes sign under time reversal and remains invariant under space inversion. The third term in Eq.~(\ref{pot}) is not invariant under rotations around the $z$ axis. The non-invariance of the potential under time reversal arises from the terms proportional to the amplitudes $f_{0e}$ and $f_{1p}$. As expected, the interaction potential matrix (\ref{pot}) satisfies the Hermiticity condition $u_{\alpha\beta}(\mathbf{k},\mathbf{k}^{'})=u_{\beta\alpha}^{*}(\mathbf{k}^{'},\mathbf{k})$, as well as the condition $\text{Im}(u_{\alpha\beta})(\mathbf{k},\mathbf{k})=0$, which ensures the absence of absorption. 
In what follows, we assume that the anisotropic correction is small compared to the isotropic terms; therefore, all terms quadratic in $f_{1e}$ and $f_{1p}$ will be omitted in the subsequent expressions.

\section{Green's Function in the Presence of Weak p-Wave Disorder} 

\label{sec:greens}

As a result of quasiparticle scattering by the aerogel strands, the disorder-averaged normal-state Green's function generally becomes dependent on the direction of the quasiparticle momentum. However, since the anisotropic part of the potential is assumed to be small, this dependence can be neglected in the limit under consideration. Indeed,
\begin{gather}
G_{\alpha\beta}(\mathbf{k},\omega_n)=\Big\{(i\omega_n-\xi_k)(\sigma_0)_{\alpha\beta}-\Sigma_{\alpha\beta}(\omega_n,\mathbf{k})\Big\}^{-1},
\end{gather}
where $\Sigma_{\alpha\beta}(\omega_n,\mathbf{k})$ is the self-energy, $\xi_{\mathbf{k}}=(k^2-k_F^2)/(2M)$ is the energy measured from the Fermi energy, $\hbar=1$, $k_F$ is the Fermi momentum, and $M$ is the particle mass. The first-order correction to the self-energy with respect to the perturbing potential describes a shift in the chemical potential:
\begin{gather}
\Sigma_{\alpha\beta}^{(1)}\cdot(2\pi)^3\delta(\mathbf{k}^{'}-\mathbf{k}) = \langle u(\mathbf{k}^{'},\mathbf{k})\rangle = (2\pi)^3\cdot n_s\times\nonumber\\
\delta(\mathbf{k}-\mathbf{k}^{'})
\Big[g_0+n_b\cdot f_{0,p}\Big](\sigma_0)_{\alpha\beta}.
\end{gather}
Here, we take into account that the ensemble average satisfies $\langle \mathbf{m}\rangle=0$. The imaginary part of the self-energy is determined from the self-consistent equation:
\begin{gather}
\Sigma_{\alpha\beta}^{(2)}(\mathbf{k})\cdot(2\pi)^3\delta(\mathbf{k}^{'}-\mathbf{k}) =\nonumber\\
\int\frac{d^3k_1}{(2\pi)^3}\frac{ \langle u_{\alpha\gamma}(\mathbf{k}^{'},\mathbf{k}_1)u_{\gamma\beta}(\mathbf{k}_1,\mathbf{k})\rangle}{i\omega_n-\xi_{\mathbf{k}_{1}}-\Sigma(\mathbf{k}_1,\omega_n)}.
\end{gather}
Writing the above expression, we have exploited the fact that both the self-energy and the Green's function are trivial with respect to the spin indices. This holds because the mean square of the potential is trivial and, within the linear approximation with respect to the anisotropic amplitude, can be readily calculated from Eq.~(\ref{pot}):
\begin{gather}
\langle u_{\alpha\gamma}(\mathbf{k}^{'},\mathbf{k}_1)u_{\gamma\beta}(\mathbf{k}_1,\mathbf{k})\rangle=(2\pi)^3\delta(\mathbf{k}-\mathbf{k}^{'})\cdot n_s\times\nonumber\\
\times\Big[\Big(g_0^2+2n_b\cdot g_0\cdot f_{0,p}\Big)\cdot\delta(k_z-(k_1)_z)+\\
n_b\cdot\Big(f_{0,p}^2+\frac{1}{3}f_{0,e}^2\Big)\label{ave_2}\Big](\sigma_0)_{\alpha\beta}.\nonumber
\end{gather}
Note that the pair correlation function entering the Dyson equation for the Green's function differs from that in the vertex-part equation and is always real-valued due to the Hermiticity of the potential. Consequently, neglecting correlations in the impurity distribution, the scalar part of the self-energy obtained from the solution of the equation takes the form:
\begin{gather}
\Sigma^{(2)}(\omega_n,\mathbf{k})=-\frac{i}{2} Mn_s\Big[g_0^2+2n_b\cdot g_0\cdot f_{0,p}
+\\n_b\frac{k_F}{\pi}\Big(f_{0,p}^2+\frac{1}{3}f_{0,e}^2\Big)
\Big]sgn(\omega_n).\nonumber
\end{gather}
The obtained expressions allow us to conclude that both the self-energy and the Green's function are isotropic. We define the inverse lifetime as follows:
\begin{gather}
\frac{1}{2\tau(\omega_n,\mathbf{k})}=i\cdot Im\big[\Sigma(\omega_n,\mathbf{k})\big],
\end{gather} 
Then, the Green's function can be rewritten in terms of standard notation:
\begin{gather}
G_{\alpha\beta}(\omega_n,\mathbf{k})=\frac{1}{i\Big[|\omega_n|+\frac{1}{2\tau}\Big]sgn(\omega_n)-\xi_k}\cdot(\sigma_0)_{\alpha\beta}.
\end{gather}
Thus, within the considered approximation, the density of states remains isotropic despite the scattering anisotropy of the model potential. Accounting for the anisotropy of the density of states, which is inherently incorporated in models of weakly anisotropic disorder ($s$-wave disorder) \cite{Aoyama, Sauls_2021, Mineev_2018}, does not affect the effect discussed below but complicates the intermediate calculations. Accordingly, the terms linear in ${k}_z$ in the potential expansion were initially omitted.

\section{Vertex Part $\Gamma_m$ and Matrix $\Lambda_{ij}$}
\label{sec:vertex}

In order to find the explicit form of the matrix $\Lambda_{jk}$ (the trivial spin structure is omitted hereafter), one needs to determine the corrections to the vertex function. In the case of anisotropic scattering, Eq.~(\ref{vertex}) is an integral equation whose solution is no longer a pure $p$-wave. Nevertheless, it can be solved analytically within the framework of the assumptions made above regarding the smallness of the anisotropic contribution. After integrating over the magnitude of the intermediate wave vector, the equation for the vertex function within the present model reduces to:
\begin{gather}
{\Gamma}_{m}(\theta,\varphi)=\hat{k}_m(\theta,\varphi)+K_1(\omega_n)\int\frac{d\Omega^{'}}{4\pi}\Gamma_m(\theta^{'},\varphi^{'})\Big\{g_{0}^2+\nonumber\\ 
2n_bg_0\Big[f_{0,p}+
if_{1,p}h_{lv}(\hat{k}_l\hat{k}_v^{'}-\hat{k}_v\hat{k}_l^{'})\Big]\Big\}\delta(\theta-\theta^{'})+\nonumber\\K_2(\omega_n)\int\frac{d\Omega^{'}}{4\pi}\Gamma_m(\theta^{'},\varphi^{'})\times\label{Gamma}\\
\times\Big[f_{0,p}^2+\frac{1}{3}f_{0,e}^2+\nonumber\\
\frac{2}{3}i\Big(f_{0,e}f_{1,e}+3f_{0,p}f_{1,p}\Big)\cdot h_{lv}\cdot(\hat{k}_l\hat{k}_v^{'}-\hat{k}_v\hat{k}_l^{'})\Big],\nonumber\\
K_1(\omega_n)=\frac{\frac{1}{2}n_s M}{|\omega_n|+\frac{1}{2\tau}},\\
K_2(\omega_n)=\frac{\frac{1}{2}n_s M\cdot \frac{n_bk_F}{\pi}}{|\omega_n|+\frac{1}{2\tau}},
\end{gather}
$h_{lm}=\langle h_{l,m}^{a,b}\rangle$ is the disorder-averaged Berry curvature tensor, which is assumed to be non-zero for a finite macroscopic region of space. Within the linear approximation with respect to the anisotropic correction, one can seek a solution by discarding higher powers of trigonometric functions (higher harmonics):
\begin{gather}
\Gamma_z(\theta,\varphi) = \gamma_{zz}^{(0)}\hat{k}_z+\gamma_{zm}^{(1)}\hat{k}_m+	\hat{k}_z^2\cdot\gamma_{zm}^{(2)}\hat{k}_m,\\
\Gamma_x(\theta,\varphi) = \gamma_{xx}^{(0)}\hat{k}_x+\gamma_{xm}^{(1)}\hat{k}_m+	\hat{k}_{\perp}^2\cdot\gamma_{xm}^{(2)}\hat{k}_m,\\
\Gamma_y(\theta,\varphi) = \gamma_{yy}^{(0)}\hat{k}_y+\gamma_{ym}^{(1)}\hat{k}_m+	\hat{k}_{\perp}^2\cdot\gamma_{ym}^{(2)}\hat{k}_m,
\end{gather}
where $\gamma_{ij}$ are angle-independent functions, $k_z^2=\cos^2\theta$, and $k_{\perp}^2=\sin^2\theta$. Solving this system yields:
\begin{gather}
\gamma_{zz}^{(0)} = \frac{1}{1 - K_1(g_0^2 + 2n_b g_0 f_{0,p})},~\gamma_{xx}^{(0)}=\gamma_{yy}^{(0)}=1,\\
\gamma_{zz}^{(1)}=\gamma_{xx}^{(1)}=\gamma_{yy}^{(1)}=0,~ \gamma_{zx}^{(1)} = -\gamma_{xz}^{(1)}=\nonumber\\
\frac{2 K_2 [f_{0,e}f_{1,e}+3f_{0,p}f_{1,p}] h_{zx}}{9 \left[1 - K_1(g_0^2 + 2n_b g_0 f_{0,p})\right]}\times i,\\
\nonumber\gamma_{xy}^{(1)} = -\gamma_{yx}^{(1)}=-i\times\frac{2}{9}K_2 [f_{0,e}f_{1,e}+3f_{0,p}f_{1,p}] h_{xy},\\
\gamma_{zz}^{(2)}=\gamma_{xx}^{(2)}=\gamma_{yy}^{(2)}=0,~\gamma_{zx}^{(2)} = -2\gamma_{xz}^{(2)}= \nonumber\\
\frac{2 K_1 n_b g_0f_{1,p} h_{zx}}{1 - K_1(g_0^2 + 2n_b g_0 f_{0,p})}\times i,\\
\nonumber\gamma_{xy}^{(2)} = -\gamma_{yx}^{(2)}=-i\times K_1 n_bg_{0}f_{1,p} h_{xy}
\end{gather}
Consequently, the coefficients $\gamma^{(1,2)}$ are non-zero only if the elements of the matrix $h_{lm}$ do not vanish. This matrix governs the simultaneous breaking of rotational and time-reversal symmetry. The primary finding of this section is that $p$-wave scattering leads to the mixing of harmonics with orbital projections differing by unity into the eigenfunction. Notably, one of the induced corrections belongs to the harmonic with $l=3$ (the coefficients $\gamma_{ij}^{(2)}$). It should be emphasized that the emergence of off-diagonal matrix elements results from interference between the conventional potential ($g_0, f_{0p}$) and exchange ($f_{0e}$) contributions to the scattering potential on the one hand, and the additional effective spin-orbit contribution ($f_{1e}, f_{1p}$) on the other. Interestingly, a similar term involving the Berry field leads to the anomalous Hall effect in a two-dimensional system of electrons moving within a matrix of ferromagnetic columns \cite{Bruno}.

We now calculate the matrix $\Lambda_{jl}$ from the Gor'kov self-consistency equation:
\begin{gather}
\Lambda_{jl}=-{\frac{3\lambda T_c}{2\pi}\cdot\sum\limits_n\int\frac{d\Omega^{'}}{4\pi}\cdot \hat{k}_j^{'}(\theta^{'},\varphi^{'})\frac{ Mk_F}{|\omega_n|+\frac{1}{2\tau(\theta^{'})}}\Gamma_l(\theta^{'},\varphi^{'})}.
\end{gather}
Using this expression and the calculated vertex function, we write down the individual matrix elements:
\begin{gather}
\Lambda_{zz}=	-\frac{3\lambda\cdot T_c M k_F}{2\pi}\sum\limits_n\int\frac{d\Omega^{'}}{4\pi}\times\nonumber\\
\times\frac{\hat{k}_z^2}{|\omega_n|+\frac{1}{2\tau(\theta^{'})}-\frac{1}{2}Mn_s\cdot \Bigl(g_0^2 + 2n_b g_0 f_{0,p}\Bigr)},\\
\Lambda_{xx}=\Lambda_{yy}=-	\frac{3\lambda\cdot T_c M k_F}{2\pi}\sum\limits_n\int\frac{d\Omega^{'}}{4\pi}\frac{\hat{k}_x^2}{|\omega_n|+\frac{1}{2\tau(\theta^{'})}},\\
\Lambda_{xz}=-\frac{3\lambda\cdot T_c M k_F}{2\pi}\sum\limits_n\int\frac{d\Omega^{'}}{4\pi}\times\nonumber\\
\times\frac{\hat{k}_x^{'}\cdot 2i\cdot n_b\cdot  h_{zx}}{|\omega_n|+\frac{1}{2\tau(\theta^{'})}-\frac{1}{2}Mn_s\cdot \Bigl(g_0^2 + 2n_b g_0 f_{0,p}\Bigr)}\times\\\nonumber
\frac{\Big[g_0f_{1,p}\cdot(\hat{k}^{'}_z)^2\hat{k}_x^{'}+\frac{[f_{0,e}f_{1,e}+3f_{0,p}f_{1,p}]k_F}{9\pi}\hat{k}_x^{'}\Big]}{|\omega_n|+\frac{1}{2\tau(\theta^{'})}},\\
\Lambda_{zx}=-\frac{3\lambda\cdot T_c M k_F}{2\pi}\sum\limits_n\int\frac{d\Omega^{'}}{4\pi}\times\nonumber\\\frac{\hat{k}_z^{'}\cdot2i\cdot n_b \cdot h_{xz}}{|\omega_n|+\frac{1}{2\tau(\theta^{'})}-\frac{1}{2}Mn_s\cdot \Bigl(g_0^2 + 2n_b g_0 f_{0,p}\Bigr)}\times\\
\times\frac{\Big[\frac{1}{2}g_0f_{1,p}\cdot(\hat{k}_{\perp}^{'})^2\hat{k}_z^{'}+\frac{[f_{0,e}f_{1,e}+3f_{0,p}f_{1,p}]k_F}{9\pi}\hat{k}_z^{'}\Big]}{|\omega_n|+\frac{1}{2\tau(\theta^{'})}}\nonumber
\end{gather}
Next, for simplicity, we consider the clean limit, i.e., we assume that $2\pi T_c^{(0)}\gg \hbar/\tau$. Integrating over the angles, summing over $n$, and expanding the digamma function, we obtain:
\begin{gather}
\Lambda_{zz}\approx-\lambda\cdot N_F\Bigl[\ln\frac{\gamma_E\cdot\omega_D}{\pi T}-\frac{Mn_sn_bk_F\Big(f_{0,p}^2+\frac{1}{3}f_{0,e}^2\Big)}{8T}\Bigr],
\end{gather}	
where $N_F = Mk_F / (2\pi^2)$ and $\gamma_E$ is the Euler constant. Since the correction to the unperturbed transition temperature $T_c^{(0)}$ is assumed to be small, we expand this expression in terms of the small deviation $\delta\tau = (T - T_c^{(0)}) / T_c^{(0)}$:
\begin{gather}
\Lambda_{zz}\approx 1+\lambda\cdot N_F\Bigl[\delta\tau+\frac{Mn_sn_bk_F\Big(f_{0,p}^2+\frac{1}{3}f_{0,e}^2\Big)}{8T_c^{(0)}}\Bigr].
\end{gather} 
In writing this expansion, we take into account that $-\lambda\cdot N_F\ln\frac{\gamma_E\cdot\omega_D}{\pi T_c^{(0)}}=1$. Similarly, for the other two diagonal elements, we obtain:
\begin{gather}
\Lambda_{xx}=\Lambda_{yy}=1+\lambda\cdot N_F\Bigl[\delta\tau+\nonumber\\
\frac{Mn_s\Big\{\pi g_0^2+2\pi n_b\cdot g_0\cdot f_{0,p}+n_bk_F\Big(f_{0,p}^2+\frac{1}{3}f_{0,e}^2\Big)\Big\}}{8T_c^{(0)}}\Bigr].
\end{gather}
For the off-diagonal elements in the clean limit, the damping-related terms in the denominators can be neglected, yielding an expression linear in the anisotropic scattering amplitude:
\begin{gather}
\Lambda_{lm} =-\Lambda_{ml} = -i\times \lambda N_F \frac{M n_s n_b}{4T_c^{(0)}} \left[ \frac{2\pi g_0f_{1,p}}{5} +\nonumber\right.\\
\frac{[f_{0,e}f_{1,e}+3f_{0,p}f_{1,p}] k_F}{9}\left.\right]\times h_{lm}
\end{gather}
Thus, owing to the symmetry breaking, the off-diagonal part of the matrix $\Lambda_{jl}$ is purely imaginary and antisymmetric, which is consistent with the Hermiticity of the full matrix. The next step is to solve the eigenvalue and eigenvector problem for the perturbed matrix.

\section{Splitting of the Superfluid Transition Temperature and the Polar-Distorted A Phase}
\label{sec:eigen}

By definition, the perturbation matrix is given by:
\begin{gather}
\delta \Lambda_{ij}=\Lambda_{ij}-\delta_{ij}.
\end{gather}
The compatibility condition for Eqs.~(\ref{Eq}), namely $\det\delta\Lambda_{ij}=0$, yields an eigenvalue equation where the eigenvalues correspond to the transition temperature shifts for different components of the order parameter. We choose the coordinate system such that $h_{zy}=0$. We assume that the system anisotropy is large, meaning that the spacing between the levels with $m=0$ and $m=\pm1$ is large. Although the spin-orbit perturbation under consideration lifts the degeneracy in the $m=\pm1$ subspace, we assume that this energy correction is small compared to the initial level splitting. In this case, perturbation theory is applicable. To write down the solution, we introduce the standard notation in terms of the mean free paths:
\begin{gather}
\frac{\pi^2}{4}\frac{\xi_0}{l_{\parallel}}=\frac{Mn_sn_bk_F\Big(f_{0,p}^2+\frac{1}{3}f_{0,e}^2\Big)}{8T_c^{(0)}},\\
\frac{\pi^2}{4}\frac{\xi_0}{l_{\perp}}=\frac{Mn_s\Big\{\pi g_0^2+2\pi n_b\cdot g_0\cdot f_{0,p}+n_bk_F\Big(f_{0,p}^2+\frac{1}{3}f_{0,e}^2\Big)\Big\}}{8T_c^{(0)}},\\
\frac{\pi^2}{4}\frac{\xi_0}{l_{xi}}=\frac{M n_s n_b |h_{xi}|}{4T_c^{(0)}} \left[ \frac{2\pi g_0f_{1,p}}{5} +\nonumber\right.\\ \left. \frac{[f_{0,e}f_{1,e}+3f_{0,p}f_{1,p} ]k_F}{9}\right],
~i=y,z,
\end{gather}
where $\xi_0 = \hbar v_F / (2\pi T_c)$ is the coherence length of superfluid $^3$He. Then, within the assumed accuracy, the corrections to the transition temperature are given by:
\begin{gather}
\delta\tau_1 = -\frac{\pi^2 \xi_0}{4l_{\parallel}}, \\
\delta\tau_2 = -\frac{\pi^2 \xi_0}{4l_{\perp}}+\frac{\pi^2 \xi_0}{4l_{xy}}, \\
\delta\tau_3 = -\frac{\pi^2 \xi_0}{4l_{\perp}}-\frac{\pi^2 \xi_0}{4l_{xy}}.
\end{gather}
In the basis under consideration, the eigenvectors take the form:
\begin{gather}
	{\mathbf{A}}_1 = \begin{pmatrix}
		-i\frac{\frac{1}{l_{xz}}}{\frac{1}{l_\perp}-\frac{1}{l_{\parallel}}} \\
		0\\
		1
	\end{pmatrix},~~ 
	{\mathbf{A}}_2 = \frac{1}{\sqrt{2}}\begin{pmatrix}
		1 \\
		i\\
		-i\frac{\frac{1}{l_{xz}}}{\frac{1}{l_\perp}-\frac{1}{l_{\parallel}}} 
	\end{pmatrix},~~\nonumber\\
	{\mathbf{A}}_3 = \frac{1}{\sqrt{2}}\begin{pmatrix}
		1 \\
		-i\\
		-i\frac{\frac{1}{l_{xz}}}{\frac{1}{l_\perp}-\frac{1}{l_{\parallel}}} 
	\end{pmatrix}.
\end{gather}
The applicability condition for the expressions above reads as follows: $\frac{1}{l_{xy}},\frac{1}{l_{xz}}\ll\frac{1}{l_\perp}-\frac{1}{l_{\parallel}}$.

The maximum transition temperature corresponds to the first solution. If $h_{xz}=0$, a pure polar phase is realized. Otherwise, the solution represents a polar-distorted A phase. 
The distortion parameter introduced at the beginning of the paper is expressed in terms of the potential parameters as:
\begin{gather}
\alpha\approx\frac{\frac{2}{l_{xz}}}{\frac{1}{l_{\perp}} - \frac{1}{l_{\parallel}}}=\nonumber\\
\frac{{2n_b |h_{xz}|}\cdot\left[ \frac{2g_0f_{1,p}}{5} + \frac{[f_{0,e} f_{1,e}+3f_{0,p} f_{1,p}]k_F}{9\pi}\right]}{g_0^2+2n_b\cdot g_0\cdot f_{0,p}}.
\end{gather}

In this section, the problem is solved within perturbation theory to demonstrate the emergence of the additional order parameter amplitude in the simplest way. However, the solution can be generalized to the case where the level spacing is small compared to the off-diagonal matrix elements, i.e., $|\Lambda_{xz}|,|\Lambda_{yz}|\gg\Lambda_{zz}-\Lambda_{xx}$. Crucially, due to the preferential alignment of the strands along the $z$ axis, the normal to the scattering surface lies in the $xy$ plane. By definition, the Berry field must be perpendicular to the scattering surface (see the Appendix or Ref.~\cite{Bruno}). It follows that after averaging over disorder, the matrix element $\langle h_{xy}\rangle$ can be neglected, and the problem reduces to the description of an ordinary two-level system. In this case, the distortion parameter is determined by:
\begin{gather}
	\tan\left(\frac{\alpha}{2}\right) = l_{xz} \left( \sqrt{\left[\frac{1}{l_{\perp}}-\frac{1}{l_{\parallel}}\right]^2+\frac{1}{l_{xz}^2}} - \left[\frac{1}{l_{\perp}}-\frac{1}{l_{\parallel}}\right] \right),
\end{gather}
which describes a continuous transition between the polar and A phases via the polar-distorted A phase as a function of the scattering parameters.

\section{Discussion}

\label{sec:results}

An analysis of the obtained expressions shows that decorating the specular aerogel strands with additional magnetic scattering centers leads to two primary effects. First, the scattering is no longer specular (the conservation law for the momentum projection along the $z$ axis is broken), which causes an additional suppression of the transition temperatures for all order parameter components and, consequently, leads to the violation of Anderson's theorem for the polar phase. In fact, a similar result was previously obtained within the model of weakly anisotropic non-magnetic impurities. Second, and this represents the main result of this work, if the scattering simultaneously breaks both axial and time-reversal symmetries, the polar phase ceases to be the solution with the maximum superfluid transition temperature. For instance, in Ref.~\cite{Mineev_2018}, the scattering potential breaks only time-reversal symmetry. Although the scattering itself is anisotropic in $k$-space, it preserves the symmetry with respect to rotations around the $z$ axis. Consequently, the polar phase remains the solution with the maximum $T_c$ in that case, even though the transition temperature is suppressed. If the axial symmetry is broken such that scattering induces transitions between states with $m=\pm1$ and $m=0$ ($p$-wave disorder), the pure polar phase ceases to exist as an eigenvector of the self-consistency equation. Instead of this state, a phase emerges with a small transverse distortion corresponding to the polar-distorted $A$ phase. We also note the well-known experimental fact that transverse compression of the aerogel does not eliminate the region where the polar phase exists \cite{Dmitriev_2019}, meaning that breaking the axial symmetry alone is likewise insufficient.

Within Ginzburg-Landau theory, the emergence of the transverse distortion can be described by writing the second-order terms in the free energy density in the form:
\begin{gather}
	\tau A_{\mu i}A_{\mu i}^{*}+\eta_{\parallel}A_{\mu z}A_{\mu z}^{*}+\eta_{\perp}\Big(A_{\mu x}A_{\mu x}^{*}+A_{\mu y}A_{\mu y}^{*}\Big)+\nonumber\\
	i\delta\eta_{ij}(\mathbf{r})(A_{\mu i}A_{\mu j}^{*}-A_{\mu i}^{*}A_{\mu j}),
\end{gather}
where $\delta\eta_{ij}$ is an antisymmetric real matrix. In the present work, we considered the case of global aerogel anisotropy, i.e., when $\langle \delta\eta_{ij}(\mathbf{r})\rangle\neq0$ (provided that $\langle \mathbf{m}\rangle=0$). As a result, the maximum transition temperature corresponds to the polar-distorted A phase, with a fixed direction of the orbital vector $\mathbf{l}$ throughout the entire volume occupied by the aerogel. In a real aerogel, the field $\delta\eta_{ij}(\mathbf{r})$ fluctuates, and its average value over the whole sample is zero. However, in this case, one should use the Larkin-Imry-Ma argument, according to which the local solution corresponds to the polar-distorted A phase, but the direction of $\mathbf{l}$ varies slowly in space over the Larkin-Imry-Ma length \cite{Volovik_1996, Volovik_2008, Fom_Sur_2017}.

From an experimental viewpoint, the main difference of the proposed mechanism for the polar-distorted A phase formation directly from the normal state is that within the Ginzburg-Landau theory applicability region, the distortion magnitude must be finite and temperature-independent. Conversely, if a narrow region where the polar phase exists is present, the distortion parameter should grow within a temperature interval of the order of this region's width until it reaches the values corresponding to the bulk A phase. In experiments \cite{Dmitriev_magnetic,Dmitriev_Eltsov}, in the case of a slight underplating of the surface with $^4$He in nafene-72 and mullite aerogel, i.e., in the presence of weak magnetic scattering, a state was observed that is likely the polar-distorted A phase with a finite distortion parameter throughout the investigated temperature range. If there is no $^4$He on the aerogel surface, the undistorted A phase is always observed. This case apparently corresponds to the limit where magnetic scattering is so strong that the regime opposite to the one considered in our problem is realized: the off-diagonal elements are much larger than the level spacing (the energies of the states with $m=0$ and $m=\pm1$ are close). This fact is consistent with the results of Ref.~\cite{Mineev_2018}, which showed that the system may become more isotropic when additional magnetic scattering is present.

The key assumption used in this work is the presence of an effective spin-orbit interaction arising from quasiparticle scattering by the magnetic aerogel strands. As noted above, this type of interaction is determined by the Berry curvature of the solid $^3$He layer magnetization texture. A qualitative derivation of this contribution to the Hamiltonian is provided in the Appendix. However, the problem of the magnetic properties of the solid layer on the aerogel surface requires a more detailed consideration from both experimental and theoretical viewpoints. At present, there are only indirect indications that the layer magnetic structure cannot be described within a simple model of independent paramagnetic centers \cite{Dmitriev_2025, Surovtsev_2025, Saunders_2021}.

Another minor but interesting finding is that the obtained expression for the vertex function contains a term with the $L=3$, $m=0,\pm1$ harmonics. This, in turn, means that the emerging phase contains a small $f$-wave component. Crucially, the number of nodes in the excitation spectrum is preserved. Interestingly, if a potential that allows transitions with a projection change of $\Delta m=\pm 2$ ($d$-wave disorder) is added to the problem, one can expect the system to exhibit four Weyl nodes on the equator.

In conclusion, we note that the investigated physical mechanism represents only one possible interpretation of the experiments by Dmitriev's group. However, the proposed mechanism for the stabilization of chiral phases could be realized in other superfluid systems with non-trivial pairing and correlated magnetic disorder.

\section*{Acknowledgments}

The author is grateful to V.~V.~Dmitriev, I.~A.~Fomin, A.~A.~Soldatov, and A.~N.~Yudin for fruitful discussions and constructive criticism.





\section{Appendix}

This appendix demonstrates that the spatial inhomogeneity of the local magnetization leads to a contribution to the scattering amplitude that is linear in the Berry curvature. A classical derivation of the Hamiltonian transformation for 2D electrons moving in a slowly varying magnetic field can be found in Ref.[21]. We consider the elastic scattering of a liquid $^{3}\text{He}$ quasiparticle by an impenetrable magnetic surface within the adiabatic approximation. The physical picture of the interaction is based on the separation of spatial scales: the barrier potential $V_0(z)$ varies on the scale of the Fermi wavelength ($\lambda_F \sim 1/k_F$), whereas the direction of the local magnetization $\mathbf{m}(\mathbf{r}_\parallel)$ varies smoothly on the macroscopic scale of the inhomogeneity or texture $R \gg \lambda_F$.

The total Hamiltonian of a quasiparticle moving in the free volume outside the aerogel strand is expressed via an effective spin-dependent surface potential of the layer:
\begin{equation}
	\hat{\mathcal{H}}_{\text{total}} = \frac{\hat{\mathbf{p}}^2}{2M} \hat{\sigma}_0 +  \hat{V}(\mathbf{r}_\parallel) \delta(z),
\end{equation}
where $z$ is the coordinate along the local normal $\mathbf{n}$, and the matrix $\hat{V}(\mathbf{r}_\parallel)$ determines the local boundary scattering amplitude at the tangential point $\mathbf{r}_\parallel$:
\begin{equation}
	\hat{V}(\mathbf{r}_\parallel)=-J\mathbf{m}(\mathbf{r}_\parallel)\boldsymbol{\sigma},
\end{equation}
where $\mathbf{m}$ is the local magnetization vector. Applying a unitary transformation $\hat{U}(\mathbf{r}_\parallel)$ such that
$\hat{U}(\mathbf{r}_\parallel)\mathbf{m}(\mathbf{r}_\parallel)\boldsymbol{\sigma}\hat{U}^{\dagger}(\mathbf{r}_\parallel)=\sigma_z$, the transformed Hamiltonian takes the form:
\begin{gather}
	\hat{\mathcal{H}}=\frac{\hat{p}_{\perp}^2\sigma_0}{2M}+\frac{(\hat{\mathbf{p}}_{\parallel}\sigma_0-\hat{\mathbf{A}}_{\parallel})^2}{2M}-J\sigma_z\delta(z),
\end{gather}
where
\begin{gather}
	\hat{{A}}_{\parallel,i}(\mathbf{r}_\parallel) = i\hbar\hat{U}^\dagger(\mathbf{r}_\parallel)\nabla_{\parallel,i}\hat{U}(\mathbf{r}_\parallel) =  \hat{{A}}_{\parallel,i}^{D} + \hat{{A}}_{\parallel,i}^{ND},
\end{gather}
where $\hat{A}_{\parallel,i}^{D} = A_{\parallel,i}^{(z)}\hat{\sigma}_z$ is the diagonal (adiabatic) part, and $\hat{A}_{\parallel,i}^{ND} = A_{\parallel,i}^{(x)}\hat{\sigma}_x + A_{\parallel,i}^{(y)}\hat{\sigma}_y$ is the off-diagonal part of the Berry potential, which describes texture-induced spin-flip transitions.
The operator $\hat{A}$ is a column of $2\times2$ matrices. By definition, the Berry potential vanishes along the surface normal ($A_\perp = 0$). If the potential varies slowly, the off-diagonal part can be neglected; thus, to first order in the gradient, the local texture only causes a phase shift of the wave functions. To construct the effective scattering Hamiltonian, we expand the operator $\hat{A}_{\parallel,j}^{D}$ near the reflection point in terms of $\mathbf{r}_{\parallel}$:
\begin{equation}
	A^D_{\parallel,j}(\mathbf{r}_{\parallel}) \approx A_{\parallel,j}^D(0) + \hat{r}_{\parallel,k} \left( \nabla_{\parallel,k} A_{\parallel,j}^D \right)_0.
\end{equation}
Choosing a gauge where $A_{\parallel,j}^D(0) = 0$, we consider the term linear in the tangential momentum $\hat{p}_{\parallel,j} = -i\hbar\nabla_{\parallel,j}$:
\begin{equation}
	\hat{\mathcal{H}}_{int} = -\frac{1}{2M} \left( \hat{p}_{\parallel,j} \hat{A}^D_{\parallel,j} + \hat{A}^D_{\parallel,j} \hat{p}_{\parallel,j} \right).
\end{equation}
Substituting the expansion and simplifying yields a term proportional to the Berry curvature:
\begin{equation}
	\hat{\mathcal{H}}_{int}^{(a)} = -\frac{1}{2M}  \left( \nabla_l A^D_{j}- \nabla_j A^D_{l}\right) \hat{r}_l \hat{p}_j\cdot\sigma_z=e_{ljk}H_k\hat{r}_l \hat{p}_j\cdot\sigma_z,
\end{equation}
where the expression for the Berry curvature follows from the definition of the gauge potential:
\begin{equation}
	H_z = \frac{\hbar}{2} \mathbf{m} \cdot \left[ \frac{\partial \mathbf{m}}{\partial x} \times \frac{\partial \mathbf{m}}{\partial y} \right] = \frac{\hbar}{2} \sin\theta \left( \frac{\partial \theta}{\partial x}\frac{\partial \varphi}{\partial y} - \frac{\partial \theta}{\partial y}\frac{\partial \varphi}{\partial x} \right),
\end{equation}
the angles $\theta$ and $\phi$ parameterize the direction of the magnetization $\mathbf{m}$. Transforming back to the laboratory frame results in the effective spin-orbit interaction used in this work:
\begin{eqnarray}
	\mathcal{H}_{eff}\sim(\mathbf{m}\cdot\boldsymbol{\sigma})\cdot(\mathbf{H}\cdot[\mathbf{r}\times\mathbf{p}]).
\end{eqnarray}

The adiabatic approximation allows for an explicit derivation of the antisymmetric tensor $h_{ij}$ associated with the Berry curvature of the magnetization texture. Since its form is determined by the symmetry of the texture, non-adiabatic processes (off-diagonal terms) beyond the adiabatic approximation are expected to primarily renormalize the coefficient of this invariant without altering its tensor structure. Therefore, the description via the Berry curvature should be understood as a qualitative microscopic justification rather than a quantitative estimation of its magnitude. The quantitative relation between $h_{ij}$ and the parameters of the magnetic texture requires a theory beyond the adiabatic limit and lies outside the scope of this work. For instance, accounting for the off-diagonal terms in the gauge potential also contributes to the phase shift of the initial-state wave function; in the second order of gradient perturbation theory, this leads to an antisymmetric fourth-order tensor with respect to the texture gradients, etc. Thus, the small expansion parameter is the ratio $\frac{\hbar^2}{2ML^2 J_0}$, where $L$ is the scale of the magnetic inhomogeneity. For $L \sim 30$~nm and $J_0 \sim 1$~mK, this parameter is of the order of $10^{-1}$.

The actual structure of the magnetic coatings on the aerogel strands is unknown. However, the presence of strand intersections, thickenings, and other inhomogeneities leads to spatial variations in the local magnetization direction both along and across the strands. Such textures possess a non-vanishing local chirality $\mathbf{m} \cdot (\nabla_i \mathbf{m} \times \nabla_j \mathbf{m})$, which within the adiabatic approximation gives the antisymmetric tensor $h_{ij}$.

\end{document}